\documentclass[a4paper,12pt]{article}

\title{
Variable selection in high-dimensional logistic regression models using a whitening approach}
\author{Wencan Zhu, C\'eline L\'evy-Leduc, Nils Tern\`{e}s}

\usepackage[nottoc]{tocbibind}
\usepackage[utf8x]{inputenc}
\usepackage{xcolor}
\usepackage{amsmath,amssymb,amsthm,amscd}
\usepackage{amsfonts}
\usepackage{graphicx}
\usepackage{rotating}
\usepackage{caption}
\usepackage{subcaption}
\usepackage{tabularx,ragged2e}
\usepackage{booktabs}
\usepackage{libertine}
\usepackage[shortlabels]{enumitem}
\usepackage{makecell}
\usepackage{multirow}
\usepackage{natbib}

\usepackage{pst-poker}
\usepackage[libertine]{newtxmath}
\usepackage{calrsfs} 
\DeclareMathAlphabet{\pazocal}{OMS}{zplm}{m}{n}
\DeclareMathAlphabet\mbi{OML}{cmm}{b}{it}
\DeclareSymbolFont{boldsymbols}{OMS}{cmsy}{b}{n}
\DeclareSymbolFontAlphabet{\mathbfcal}{boldsymbols}

\DeclareMathOperator*{\argmax}{arg\,max}
\DeclareMathOperator*{\argmin}{arg\,min}

\usepackage{array}
\newcommand{\by}{\mathbf{y}}
\newcommand{\bX}{\mathbf{X}}

\newcommand{\bw}{\mathbf{w}}

\newcommand{\bH}{\mathbf{H}}

\newcommand{\bbeta}{\boldsymbol{\beta}}

\newcommand{\bSigma}{\boldsymbol{\Sigma}}
\newcommand{\norm}[1]{\left\lVert#1\right\rVert}

\everymath{\displaystyle}

\theoremstyle{definition}

\theoremstyle{plain}

\graphicspath{{graphics/}}

\definecolor{blue1}{HTML}{4865A0}
\definecolor{blue2}{HTML}{4C8EBD}
\definecolor{blue3}{HTML}{4FAAD1}

\begin{document}
\maketitle

\begin{abstract}
In bioinformatics, the rapid development of sequencing technology has enabled us to collect an increasing amount of omics data. Classification based on omics data is one of the central problems in biomedical research. However, omics data usually has a limited sample size but high feature dimensions, and it is assumed that only a few features (biomarkers) are active, i.e. informative to discriminate between different categories (cancer subtypes, responder/non-responder to treatment, for example). Identifying active biomarkers for classification has therefore become fundamental for omics data analysis. Focusing on binary classification, we propose an innovative feature selection method aiming at dealing with the high correlations between the biomarkers. Various research has shown the notorious influence of correlated biomarkers and the difficulty of accurately identifying active ones. Our method, WLogit, consists in whitening the design matrix to remove the correlations between biomarkers, then using a penalized criterion adapted to the logistic regression model to select features. The performance of WLogit is assessed using synthetic data in several scenarios and compared with other approaches. The results suggest that WLogit can identify almost all active biomarkers even in the cases where the biomarkers are highly correlated, while the other methods fail, which consequently leads to higher classification accuracy. The performance is also evaluated on the classification of two Lymphoma subtypes, and the obtained classifier also outperformed other methods. Our method is implemented in the \texttt{WLogit} R package available from the Comprehensive R Archive Network (CRAN). 
\end{abstract}

\section{Introduction}
With the advances in high-throughput molecular techniques, omics technologies can generate large-scale molecular data, such as genomic, transcriptomic, proteomic, and metabolomic data. Classification based on the molecular levels is one of the essential issues in genome research. Examples include tumor classification \citep{Quackenbush2006}, disease classification \citep{Loscalzo2007} and distinguishing between responder v.s. non-responder to a treatment \citep{Gustafsson2014}. \textcolor{black}{Different machine learning techniques have been applied to solve this classification problem.} Compared to classifiers such as decision tree \citep{Utgoff1989} and SVM \citep{Cortes1995}, logistic regression \citep{Walker1967} is a popular classification method with an explicit statistical interpretation and can provide classification probabilities for a binary response \citep{Menard2002}. \\

However, classification based on omics data is a challenging task. In most omics datasets, the number of biomarkers is much larger than the sample size. Under such a situation, it is generally believed that only a few biomarkers are relevant to disease outcomes, they are called active biomarkers. The presence of irrelevant biomarkers can lead to overparameterized models that increase the risk of overfitting \citep{Sung2012}. Therefore, selecting the active biomarkers can simplify the classifier without the loss of classification accuracy and ease the computational burden. Various methods for feature selection in bioinformatics were developed, and reviews can be found in \cite{Ang2015} and \cite{Jardilier2018}.
To address this issue, regularization via the Lasso \citep{Tibshirani1996} is often implemented to reduce the subset of biomarkers. It adds a penalty equal to the sum of the absolute value of the coefficients that can result in sparse models with few non-zero coefficients and eliminate biomarkers with zero coefficients.\\

To formally state the statistical problem, given a design matrix $\bX$ of size $n\times p$, $X_{j}^{(i)}$ corresponds to the measurement of the $j$th biomarker for the $i$th sample, and $\bbeta=(\beta_{1},\ldots, \beta_p)^{T}$ is the vector of effect size for each biomarker, with most components equal to zero. We assume that the binary responses $y_1,y_2,...,y_n$ are independent random variables having a Bernoulli distribution with parameter $\pi_{\bbeta}(X^{(i)})$ ($y_{i} \sim Bernoulli(\pi_{\bbeta}(X^{(i)}))$), 
where for all $i$ in $\{1,\dots,n\}$,
\begin{equation}\label{eq:logistic}
    \pi_{\bbeta}(X^{(i)})=\frac{\exp\left({\sum_{j=1}^p \beta_j X_j^{(i)}}\right)}{1+\exp\left({\sum_{j=1}^p \beta_j X_j^{(i)}}\right)}.
\end{equation}

The logistic regression with $\ell_1$ regularization solves the feature selection problem by adding a penalty function to the log-likelihood of the logistic regression model:
\begin{equation}\label{eq:lasso_logistic}
\widehat{\bbeta}= \argmin_{\bbeta} \left\{l(\bbeta) +\lambda\norm{\bbeta}_{1}\right\},
\end{equation}
where $\norm{\bbeta}_{1}=\sum_{k=1}^p |\beta_k|$, and the log-likelihood $l(\bbeta)$ is defined by:
\begin{equation}\label{eq:ll_logistic}
l(\bbeta) = \frac{1}{n}\sum_{i=1}^{n}\left[y_{i}\cdot X^{(i)} \bbeta-\log(1+e^{X^{(i)}\bbeta})\right],
\end{equation}
with $X^{(i)}$ the $i$th row of $\bX$. With the penalty function and properly chosen parameter $\lambda$, some components of $\widehat{\bbeta}$ are set to zero. 
Recently, penalization approaches have been widely applied to biomarker discovery and disease classification \citep{Zhu2004, Wu2006, Ma2008, Liu2020}. A more comprehensive review of different regularizations for analyzing high-dimensional omics data can be found in \cite{Vinga2021}. \\

Despite various advantages, the Lasso criterion can fail to select the true subset of active biomarkers when all biomarkers are highly correlated, especially when the correlation between active and non-active biomarkers is large. This phenomenon was explicitly explained by \cite{Zhao2006}, where a condition is established for Lasso to consistently select the true model in the classical Gaussian regression model. The condition is called the Irrepresentable Condition (IC) (or incoherent condition by \cite{Meinshausen2009}), and related properties in a Gaussian linear model were reached independently by \cite{Zhao2006} and \cite{Meinshausen2009}. A similar condition was obtained by \cite{Ravikumar2010} and \cite{Bunea2008} in the logistic regression case. Let Q be defined by:
\begin{equation}
    Q=\bX^{T}\bH\bX,
\end{equation}
where $\bH$ is a diagonal matrix with 
\begin{equation}\label{eq:H}
H_{ii}=\pi_{\bbeta}(X^{(i)})/(1-\pi_{\bbeta}(X^{(i)})), 1 \leq i\leq n.
\end{equation}

Let $S=\{j,\;\beta_{j}\neq0\}$ be the set of active variables with size $d$, $S^{c}$ the set of non-active variables. $Q_{SS}$ denotes the $d\times d$ sub-matrix of $Q$ indexed by $S$. With this notation, the condition states:

There exists $\alpha\in(0,1]$ such that:
\begin{equation}\label{eq:LIC}
     \left|Q_{S^{c}S}(Q_{SS})^{-1} \right|_{\infty}\leq 1-\alpha, 
\end{equation}
where  $\left|A\right|_{\infty}=\max_{j=1,\ldots,p}\sum_{k=1}^{p}|A_{jk}|$ for any real symmetric matrix having $p$ rows and $p$ columns. 

To deal with the correlations between variables, several methods have been proposed. The most well-known ones include Elastic Net \citep{Zou2005} and Adaptive Lasso \citep{Zou2006}. The former combines the $\ell_1$ and $\ell_2$ penalties, and the latter assigns weights to each of the parameters in forming the $\ell_1$ penalty of Lasso.  
Several filter approaches were also proposed to take into consideration the correlations in the classification framework. Relief \citep{Kira1992} is sensitive to feature interactions and has inspired a family of Relief-based feature selection algorithms, notably the ReliefF \citep{Kononenko1997}. It was widely used in biomedical research \citep{Urbanowicz2018}. Fast Correlation Based Filter (FCBF) \citep{Yu2003} is another approach in high-dimensional feature selection that evaluates feature relevance and redundancy based on correlation measures.\\


In this article, we propose a novel feature selection method to take this issue into account by removing the correlations between biomarkers in the high dimensional logistic regression model. Inspired by the idea of WLasso (Whitening Lasso) proposed by \cite{Zhu2021}, we first ‘whiten’ the columns of $\bX$. Then, the biomarker selection is performed thanks to a regularized quadratic approximation of the log-likelihood. More details on this method are presented in Section 2. In Section 3, the performance of the proposed method is assessed via numerical experiments and compared with several methods focusing on the same problem. In Section 4, we apply the proposed procedure to a publicly available omic dataset aiming at identifying active biomarkers to classify on two Lymphoma subtypes. Finally, we discuss our findings and give concluding remarks in Section 5.

\section{Method}
To solve the optimization problem (\ref{eq:lasso_logistic}), one may directly minimize the penalized log-likelihood \citep{Park2007, Wang2019}, or use least square approximation as proposed by \cite{Friedman2010}, 
which proposes to form a quadratic approximation of the log-likelihood (\ref{eq:ll_logistic}) by using a Taylor expansion at the current estimates:

\begin{align}\label{eq:newton_glmnet}
     l_{Q}(\bbeta)&=-\frac{1}{2n}\sum_{i=1}^{n}w_{i}(z_{i}-X^{(i)}\bbeta)^{2}+C(\bbeta^{o})^{2}\\
    &=-\frac{1}{2n}\sum_{i=1}^{n}(\sqrt{w_{i}}z_{i}-\sqrt{w_{i}}X^{(i)}\bbeta)^{2}+C(\bbeta^{o})^{2}
\end{align}
with
\begin{equation*}
    z_{i}=X^{(i)}\bbeta+\frac{y_{i}- \pi_{\bbeta^{o}}(X^{(i)})}{ \pi_{\bbeta^{o}}(X^{(i)})(1- \pi_{\bbeta^{o}}(X^{(i)}))}, \textrm{(working response)}
\end{equation*}

\begin{equation}\label{eq:w}
    w_{i}= \pi_{\bbeta^{o}}(X^{(i)})(1- \pi_{\bbeta^{o}}(X^{(i)})), \textrm{(weights)}
\end{equation}
where $\pi_{\bbeta^{o}}(X^{(i)})$ is the evaluation of $\pi_{\bbeta}$ (defined in Model (\ref{eq:logistic})) at the current parameters $\bbeta^{o}$. The final estimator can be derived by the IRLS (Iterative Re-weighted Least Square) algorithm \citep{Daubechies2010}. \\

Interestingly, the logistic irrepresentable condition (\ref{eq:LIC}) coincides with the Irrepresentable condition in linear regression \citep{Zhao2006}, when replacing the matrix $\bX$ by $\sqrt{\bw}\bX$, where $\sqrt{\bw}$ is a diagonal matrix with diagonal entries equal to $(\sqrt{w_{1}}, \ldots, \sqrt{w_{n}})$ as defined in (\ref{eq:w}).

\subsection{Transformation}\label{sec:transformation}
\begin{figure}[h]
   \centering
    \includegraphics[width=0.9\textwidth]{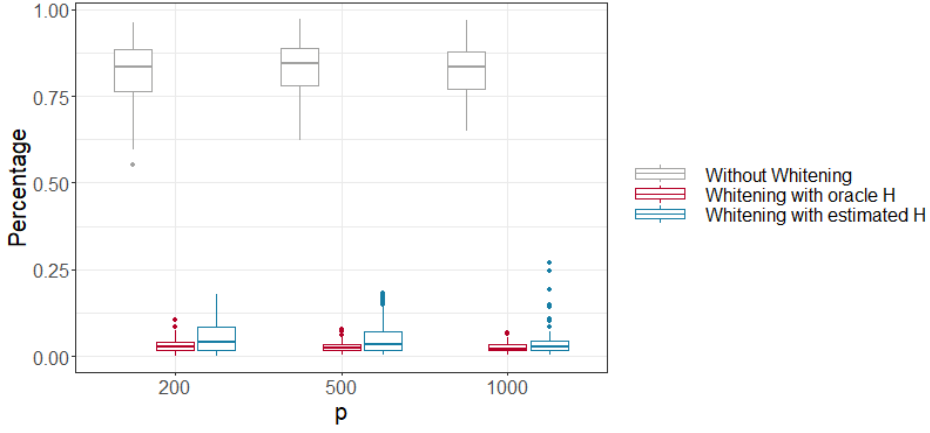}
    \caption{Percentage of elements on the left hand-side of Equation (\ref{eq:LIC}) that violated the IC, before and after transformation with oracle $\bH$ and estimated $\bH$.}
    \label{fig:IC_compare}
\end{figure}

Since the inconsistency of the Lasso estimator comes from the correlations between the biomarkers, we propose to remove the correlation by ''whitening'' the matrix $\bX$. More precisely, we consider $\widetilde{\bX}=\bX\Check{\bSigma}^{-1/2}$, where $\Check{\bSigma}$ is an covariance estimator obtained from $\bH^{1/2}\bX$ where $\bH$ is defined in Equation (\ref{eq:H}). With this transformation, $\widetilde{\bX}^{T}\bH\widetilde{\bX}$ should be close to the identity matrix $I_{p}$, thus the irrepresentable condition should be satisfied. Figure \ref{fig:IC_compare} shows the percentage of elements on the left-hand side of Equation (\ref{eq:LIC}) that violated the condition. \textcolor{black}{Data for illustration was generated on one scenario in numerical experiments: the balanced case with blockwise correlation structure when $p=500$. This dataset will be used in the rest of the section to illustrate different steps in our method.} Since in practice we do not know $\pi_{\bbeta}(X^{(i)})$, the oracle $\bH$ with true coefficients and estimated $\bH$ (see Section \ref{sec:Sigma_est} for details) were both presented. We verified through this figure that the violation of the irrepresentable condition had been reduced after the transformation.

After the whitening step, Model (\ref{eq:logistic}) can be rewritten as:
\begin{equation}\label{eq:new_logistic}
    \pi_{\widetilde{\bbeta}}(\widetilde{X}^{(i)})=\frac{\exp\left({\sum_{j=1}^p \widetilde{\beta}_j \widetilde{X}_j^{(i)}}\right)}{1+\exp\left({\sum_{j=1}^p \widetilde{\beta}_j \widetilde{X}_j^{(i)}}\right)},
\end{equation}
where $\widetilde{X}^{(i)}$ denotes the $i$th row of $\widetilde{X}$, and $\widetilde{\bbeta}=\Check{\bSigma}^{1/2}\bbeta$. 
The log-likelihood after the transformation can be written as:
\begin{equation}\label{eq:ll_new}
    l^{wt}(\widetilde{\bbeta})=\frac{1}{n}\sum_{i=1}^{n}\left\{y_{i}\cdot\widetilde{X}^{(i)}\widetilde{\bbeta}-\log\left(1+e^{\widetilde{X}^{(i)}\widetilde{\bbeta}}\right)\right\}.
\end{equation}
 Following the same technique of approximation as in (\ref{eq:newton_glmnet}), we can form a quadratic approximation to the transformed (whitened) log-likelihood (\ref{eq:ll_new}), then an estimator of $\widetilde{\bbeta}$ is obtained by solving the following problem:
\begin{equation}\label{eq:weighted_ls_new}
    \argmin_{\widetilde{\bbeta}\in \mathbb{R}^{p}} \left\{l_{Q}^{wt}(\widetilde{\bbeta})+\lambda\norm{\Check{\bSigma}^{-1/2}\widetilde{\bbeta}}_{1}\right\}.
\end{equation}

\subsection{Estimation of $\widetilde{\bbeta}$}\label{sec:est_beta_tilde}
The estimation is obtained by using an iterative procedure. Let $maxit$ and $tol$ denote the maximum number of iterations and the tolerance. For a fixed $\lambda$, the following loops are performed: 
\begin{itemize}
        \item Initialize parameters $\widetilde{\bbeta}^{(0)}$  by $\widetilde{\bbeta}^{(0)}=\Check{\bSigma}^{1/2}\bbeta^{(0)}$, where $\bbeta^{(0)}$ is obtained by ridge regression in the logistic regression model.
        \item For iteration $j=1,\ldots,maxit$:
         \begin{enumerate}
             \item Update working response, weights, weighted response, weighted design matrix in the re-weighted least square regression. 
             \item Update coefficients $\widehat{\widetilde{\bbeta}}^{(j)}$ by solving Equation (\ref{eq:weighted_ls_new}).
             \item Calculate $max(|\widehat{\widetilde{\bbeta}}^{(j)}-\widehat{\widetilde{\bbeta}}^{(j-1)}|)$
             \item For $j>1$, if $max(|\widehat{\widetilde{\bbeta}}^{(j)}-\widehat{\widetilde{\bbeta}}^{(j-1)}|)<tol$, stop and return $\widehat{\widetilde{\bbeta}}^{(j-1)}$. If $j=maxit$, stop the algorithm and return $\widehat{\widetilde{\bbeta}}^{(j)}$. If none of these conditions is satisfied, go back to Step 1 until one of the stopping criteria is satisfied.
         \end{enumerate}
        \item Denote the final coefficients by $\widehat{\widetilde{\bbeta}}_{0}(\lambda)$.
\end{itemize}

To estimate $\widetilde{\bbeta}$, we will not directly use $\widehat{\widetilde{\bbeta}}_{0}(\lambda)$ but the following modified estimator which can be seen as a correction of the components of $\widehat{\widetilde{\bbeta}}_{0}(\lambda)$.
For $K$ in $\{1,\ldots,p\}$, let $\textrm{Top}_K$ be the set of indices corresponding to the $K$ largest values of the components of $|\widehat{\widetilde{\bbeta}}_{0}|$, then the estimator of $\widetilde{\bbeta}$ is $\widehat{\widetilde{\bbeta}}=(\widehat{\widetilde{\bbeta}}_j^{(\widehat{K})})_{1\leq j\leq p}$, where $\widehat{\widetilde{\bbeta}}_j^{(K)}$ is defined by:
\begin{equation}\label{eq:beta_tilde_thresh}
 \widehat{\widetilde{\bbeta}}_{j}^{(K)}(\lambda)=
	\begin{cases}
	  \widehat{\widetilde{\bbeta}}_{0j}(\lambda), & j \in \textrm{Top}_K \\
	 \textrm{$K$th largest value of } |\widehat{\widetilde{\bbeta}}_{0j}| , & j \not\in \textrm{Top}_K.
\end{cases}
\end{equation}

To choose the parameter $K$, we use a strategy based on the log-likelihood of the model. By replacing $\widetilde{\beta}$ in (\ref{eq:ll_new}) by $\widehat{\widetilde{\beta}}^{(K)}(\lambda)$, which is the vector having the ${\widehat{\widetilde{\beta}}_j}^{(K)}$
for components, we get $l^{wt}_{K}(\widehat{\widetilde{\bbeta}}(\lambda))$, and $\widehat{K}$ is chosen as follows

$$\widehat{K}(\lambda)=\argmin \left\{K\geq 1 \textrm{ s.t. } \frac{l^{wt}_{K}(\widehat{\widetilde{\bbeta}}(\lambda))}{l^{wt}_{K+1}(\widehat{\widetilde{\bbeta}}(\lambda))}\geq\gamma\right\},\textrm{ where } \gamma\in (0,1).$$

\begin{figure}[h]
    \centering
    \includegraphics[width=0.9\textwidth]{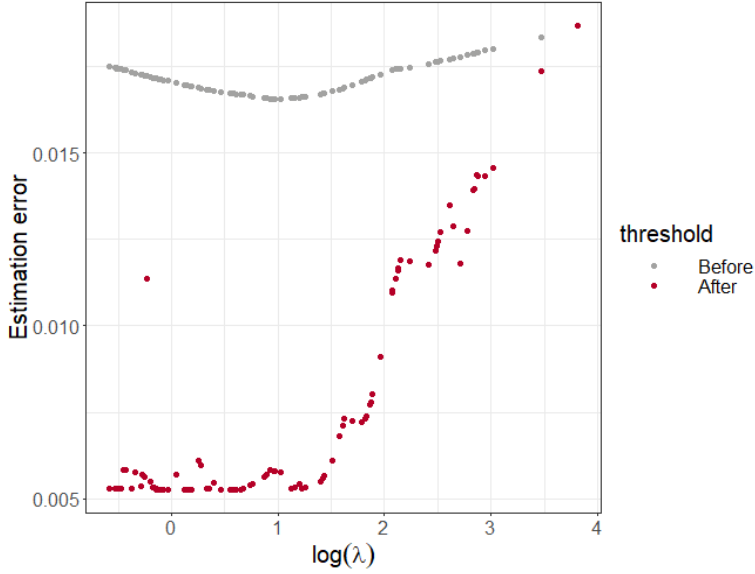}
    \caption{Average estimation error for all coefficients before and after the thresholding.}
    \label{fig:thresh_tilde}
\end{figure}

The purpose of this step is to correct the intermediate estimation $\widehat{\widetilde{\bbeta}}$. Figure \ref{fig:thresh_tilde} displays coefficient estimation error of $\widetilde{\bbeta}$ before and after the thresholding correction. We can see that the correction helps to decrease the coefficient estimation error. 

\subsection{Estimation of $\bbeta$}\label{sec:est_beta}
Resulting from the transformation, a first estimation of $\bbeta$ is obtained by $\widehat{\bbeta}_0=\Check{\bSigma}^{-1/2}\widehat{\widetilde{\bbeta}}$, and we apply a threshold to get the final estimation $(\widehat{\bbeta}_j^{(\widehat{M})})_{1\leq j\leq p}$ where

\begin{equation}\label{eq:beta_hat}
\widehat{\bbeta}_{j}^{(M)}(\lambda)=
 \begin{cases}
	  \widehat{\bbeta}_{0j}(\lambda), & j \in \textrm{Top}_M \\	  0 , & j \not\in \textrm{Top}_M,
 \end{cases}
\end{equation}
and $Top_M$ is defined in a similar way as previously. The choice of the parameter $M$ was also based on the log-likelihood. By replacing $\beta$ in (\ref{eq:ll_logistic}) by $\widehat{\beta}^{(M)}(\lambda)$, which is the vector having the
${\widehat{\beta}_j}^{(M)}$ for components, we get $l_{M}(\widehat{\bbeta}(\lambda))$. 
Using the same strategy as in Section \ref{sec:est_beta_tilde}, $M$ is chosen as follows:
$$\widehat{M}(\lambda)=\argmin \left\{K\geq 1 \textrm{ s.t. } \frac{l_{M}(\widehat{\bbeta}(\lambda))}{l_{M+1}(\widehat{\bbeta}(\lambda))}\geq\gamma\right\},\textrm{ where } \gamma\in (0,1).$$



As we can see from Figure \ref{fig:thresh_final} in Supplementary, the thresholding step successfully removed non active variables while keeping most of the true active ones in the model.

\subsection{Choice of the parameter $\lambda$}\label{sec:lambda}
Suppose the estimation of $\bbeta$ was obtained following Section \ref{sec:est_beta_tilde} and Section \ref{sec:est_beta}. For simplicity, we note it as $\widehat{\bbeta}(\lambda)$ over the sequence of $\lambda$, and the corresponding log-likelihood is $l(\widehat{\bbeta}(\lambda))$. We chose $\lambda$ by:
\begin{equation}
    \hat{\lambda}= \argmax_{\lambda}l(\widehat{\bbeta}(\lambda)).
\end{equation}
Notice that if multiple $\lambda$s maximize the log-likelihood, we chose the one leading to the most parsimonious model. 




\subsection{Estimation of $\Check{\bSigma}$}\label{sec:Sigma_est}
In practice, $\Check{\bSigma}$ is calculated by estimating the variance-covariance matrix of $\bH^{1/2}\bX$. As the diagonal of $\bH$ defined in Equation (\ref{eq:H}) is unknown 
because no information on $\bbeta$ is available, the latter can be roughly estimated by ridge regression in the logistic regression model when $p>n$. 
We denote this estimator by $\widehat{\bbeta}_{ridge}$ and obtain $\widehat{\bH}$ with $\widehat{H_{ii}}=\pi_{\widehat{\bbeta}_{ridge}}(X^{(i)})/(1-\pi_{\widehat{\bbeta}_{ridge}}(X^{(i)}))$ for $i=1,\ldots,n$. Finally, $\Check{\bSigma}$ is calculated by estimating the variance-covariance matrix from $\widehat{\bH}^{1/2}\bX$, by using the method implemented in the package cvCovEst of \cite{Boileau2022}.

\subsection{Summary of WLogit algorithm}

\begin{enumerate}
    \item Calculate $\Check{\bSigma}$, the empirical variance-covariance matrix of $\bH^{1/2}\bX$, as described in Section \ref{sec:Sigma_est}
    \item Compute $\widetilde{\bX}=\bX\Check{\bSigma}^{-1/2}$
    \item For each $\lambda$:
     \begin{enumerate}
         \item Estimate $\widetilde{\bbeta}$ as described in Section \ref{sec:est_beta_tilde}.
         \item Estimate $\bbeta$ as described in Section \ref{sec:est_beta}.
     \end{enumerate}
    \item Choose $\lambda$ as described in Section \ref{sec:lambda}, then perform variable selection and/or prediction of $\by$ based on $\widehat{\bbeta}(\hat{\lambda})$.
\end{enumerate}

\section{Numerical experiments}
This section aims at evaluating WLogit and comparing it with other existing methods. We simulated data from Model (1), where the rows of $\bX$ are assumed to be independent Gaussian random vectors with covariance matrix equal to $\bSigma$. The response $\by$ was generated following Model (1), and the vector $\bbeta$ has 10 non-zero elements with an effect size equal to 1. The sample size is equal to $n=100$, and we considered the balanced case where there are 50 responses $y_i$ equal to 1 and 50 equal to 0, and an imbalanced case where there are 20 responses $y_i$ equal to 1 and 80 equal to 0. The number of predictors (biomarkers) took its values from 200 to 2000. 100 replications were generated for each scenario.\\

In our simulations, we mainly considered correlation structures in which the irrepresentable condition was violated. We defined $\bSigma$ with a blockwise structure:
\begin{equation}\label{Sigma_structure}
    \bSigma=
\begin{bmatrix}
	   \bSigma_{11} &  \bSigma_{12} \\
	    \bSigma_{21} &  \bSigma_{22} 
\end{bmatrix},
\end{equation}
where $\bSigma_{11}$ (resp. $\bSigma_{22}$) are the correlation matrix of active (resp. non-active) biomarkers with off-diagonal entries equal to $\alpha_{1}$ (resp. $\alpha_{3}$), $\bSigma_{12}$ is the correlation matrix between active and non-active variables with entries equal to $\alpha_{2}$. In our simulations, we chose $(\alpha_{1}, \alpha_{2}, \alpha_{3}) = (0.3, 0.5, 0.7)$, one of the frameworks proposed by \cite{Xue2017}. Although this structure was proposed in the context of linear regression, we checked that the irrepresentable condition for the logistic model was also violated (as displayed in Figure 1). Additionally to this special case, we also investigated the case where no correlation exists between predictors, i.e., $\bSigma$ is the identity matrix, and in this case, the irrepresentable condition is satisfied.\\

\subsection{Compared methods}
Compared methods include two other penalized approaches: Lasso and Elastic Net adapted to the logistic regression model. Elastic Net is noted as EN in the figures. The parameters in these two algorithms are chosen by 10-fold cross-validation and implemented by the R package \texttt{glmnet}. We also compared our method with other approaches not involving the penalized regression family: ReliefF and FCBF. They also take into account the correlations between predictors and are widely used in the identification of biomarkers. ReliefF was implemented by the R package \texttt{CORElearn} with parameter \texttt{estimator="ReliefFexpRank"}. Since this method only gives the rank of predictors, we selected the same number of predictors as WLogit with the highest rank. FCBF was implemented by the Bioconductor package \texttt{FCBF}. We kept the default parameters for these two methods.

\subsection{Evaluation}
\begin{figure}[h]
    \centering
    \includegraphics[width=0.99\textwidth]{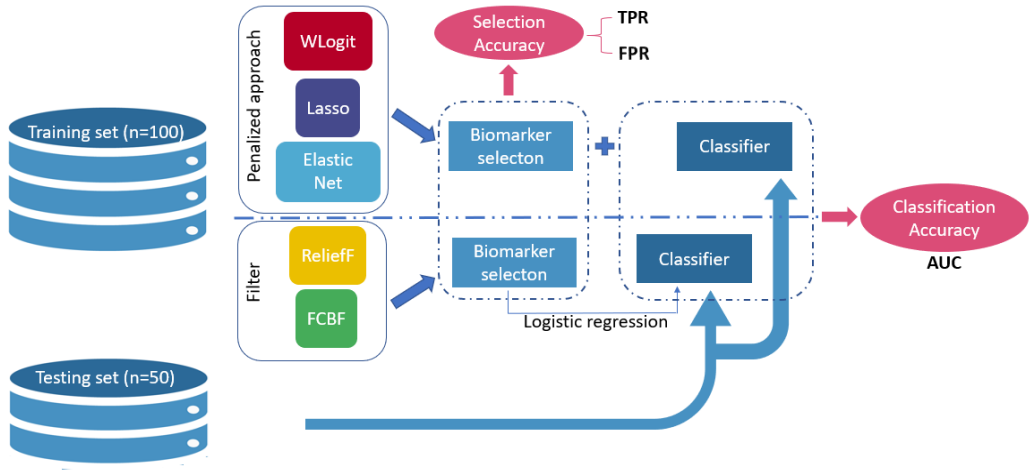}
    \caption{Simulation process and evaluation of the compared methods. }
    \label{fig:simulation}
\end{figure}

The evaluation of the performance of the compared methods was based on two aspects: (1) the accuracy of biomarker selection and (2) the accuracy of sample classification, which can be seen as a prediction task. Figure \ref{fig:simulation} shows different steps in the numerical experiments and the two types of evaluation. \\

\subsubsection*{Biomarker selection}
We generate training sets as described at the beginning of this section. Each method selected a subset of predictors, and we evaluate the selection by True Positive Rate (TPR) and False Positive Rate (FPR). The reported values for TPR and FPR are obtained by averaging these values from 100 replications.

\subsubsection*{Sample classification}
For penalized regression approaches (WLogit, Lasso, and Elastic Net), a classifier was already available with selected predictors since these approaches also give regression coefficients estimation at the same time. For ReliefF and FCBF, when a subset of predictors was chosen, the logistic regression classifier was built with the estimation of coefficients on each chosen predictor. The evaluation was then performed on another simulated testing set with the same settings as the training set, except with only half the sample size (100 (training) v.s. 50 (testing)). The evaluation on the testing test will provide the prediction accuracy of the selected set of predictors, which is presented by the AUC (Area Under the receiver operating characteristic (ROC) curve).

\subsection{Results}
\begin{figure}[h]
    \centering
    \includegraphics[width=0.9\textwidth]{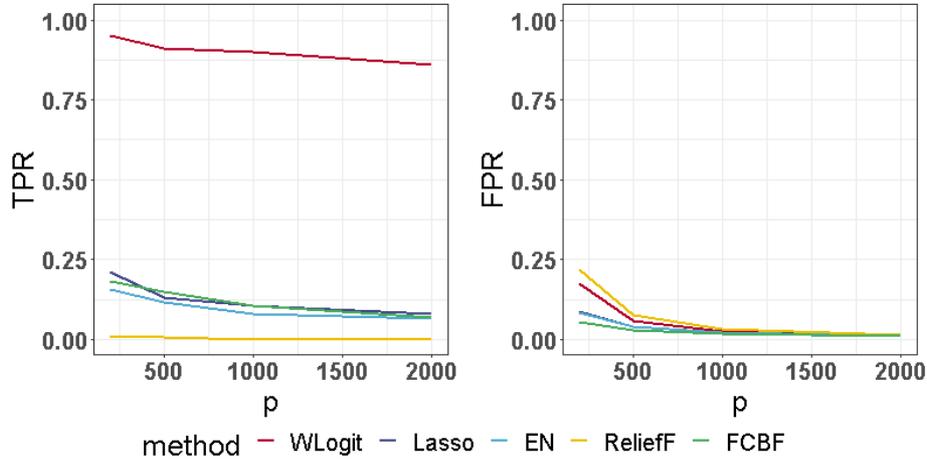}
    \caption{True Positive Rate (left) and False Positive Rate (right) for different methods in the balanced case when $\bSigma$ is defined in (\ref{Sigma_structure}) with $(\alpha_1,\alpha_2, \alpha_3)= (0.3, 0.5, 0.7)$.}
    \label{fig:comp_TF_357_1}
\end{figure}

\begin{figure}[h]
    \centering
    \includegraphics[width=0.9\textwidth]{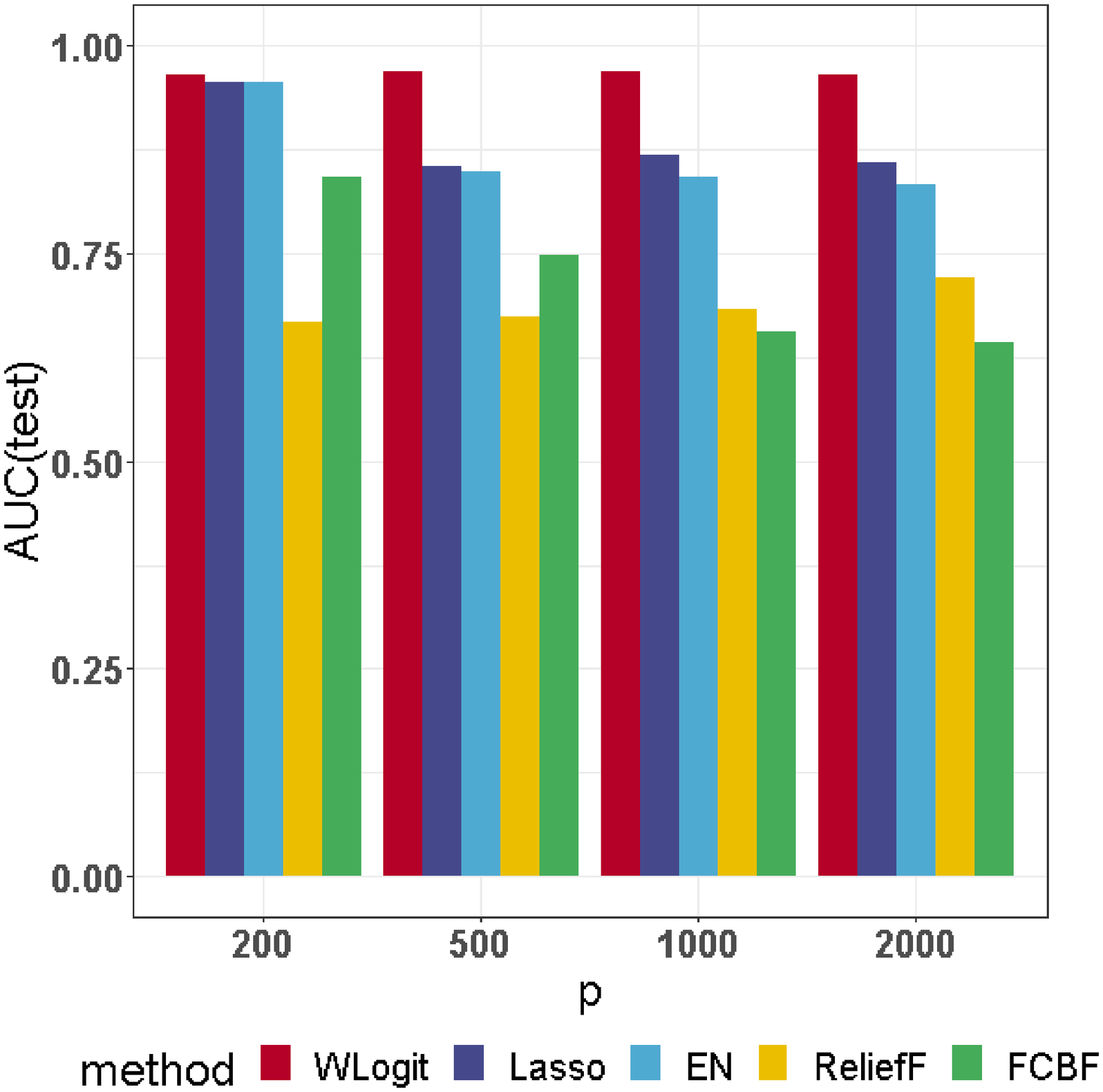}
    \caption{AUC on the testing set for different methods in the balanced case when $\bSigma$ is defined in (\ref{Sigma_structure}) with $(\alpha_1,\alpha_2, \alpha_3)= (0.3, 0.5, 0.7)$.}
    \label{fig:AUC_357_1}
\end{figure}

The corresponding results are displayed in Figures \ref{fig:comp_TF_357_1} and \ref{fig:AUC_357_1} in the case where $\bSigma$ has the blockwise correlation structure defined in Model (\ref{Sigma_structure}) with parameters $(\alpha_1,\alpha_2, \alpha_3)= (0.3, 0.5, 0.7)$. The corresponding TPR and FPR for each method are displayed. We can see from Figure \ref{fig:comp_TF_357_1} that WLogit largely outperforms the other methods: the TPR is always the largest and close to 1 (0.95 for $p=200$ and 0.86 for $p=2000$ ). Lasso, Elastic Net, and FCBF performed similarly. They can identify a very limited number of active variables (TPR smaller than 0.20). Although the FPR for WLogit was larger when $p=200$ (FPR$=0.17$), it decreased when $p$ increases (FPR$=0.01$ for $p=2000$). When $p=2000$, the FPR for all the methods is similar. With the same subset size of selected variables as WLogit, ReliefF performed poorly: the TPR is close to 0, and the FPR is the largest when $p$ is not large. \\

Figure \ref{fig:AUC_357_1} presents the average of AUC on the testing set for all methods, based on the classifiers developed on the training set (variable selection evaluated in Figure \ref{fig:comp_TF_357_1}). WLogit showed the best classification accuracy stable at a high level ($>0.96$) even when the number of predictors increases, which may come from the fact that it has identified more active variables than others. Lasso and Elastic Net performed similarly (AUC$=0.86$ and $0.83$ for Lasso and Elastic Net, respectively, when $p=2000$). Although FCBF showed competitive predictor selection accuracy, the classification accuracy (AUC$=0.64$ when $p=2000$) is lower than the one of Lasso and Elastic Net. Moreover, its classification accuracy decreased with the increase of $p$ and was even lower than Relief from $p=1000$ (0.65 for FCBF and 0.68 for Relief when $p=1000$). \textcolor{black}{This may come from the fact that the selected biomarkers from FCBF underwent a re-estimations of coefficients by a logistic regression, while for Lasso and Elastic Net, their coefficients were directly derived from the feature selection step, which provided more accurate prediction.} \\

\begin{figure}[h]
    \centering
    \includegraphics[width=0.9\textwidth]{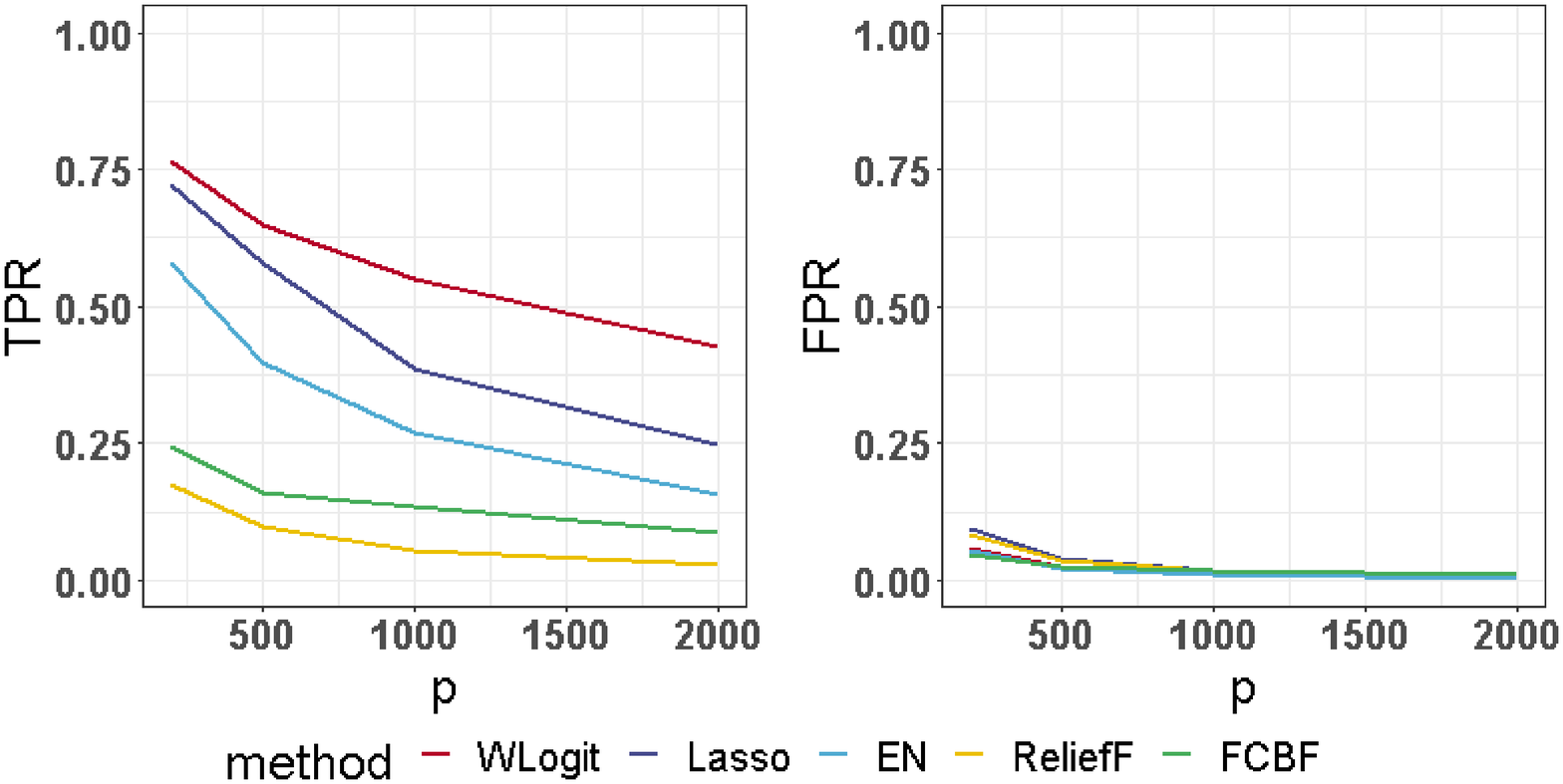}
    \caption{True Positive Rate (left) and False Positive Rate (right) for different methods in the balanced case when $\bSigma$ is the identity matrix.}
    \label{fig:comp_TF_000_1}
\end{figure}

\begin{figure}[h]
    \centering
    \includegraphics[width=0.9\textwidth]{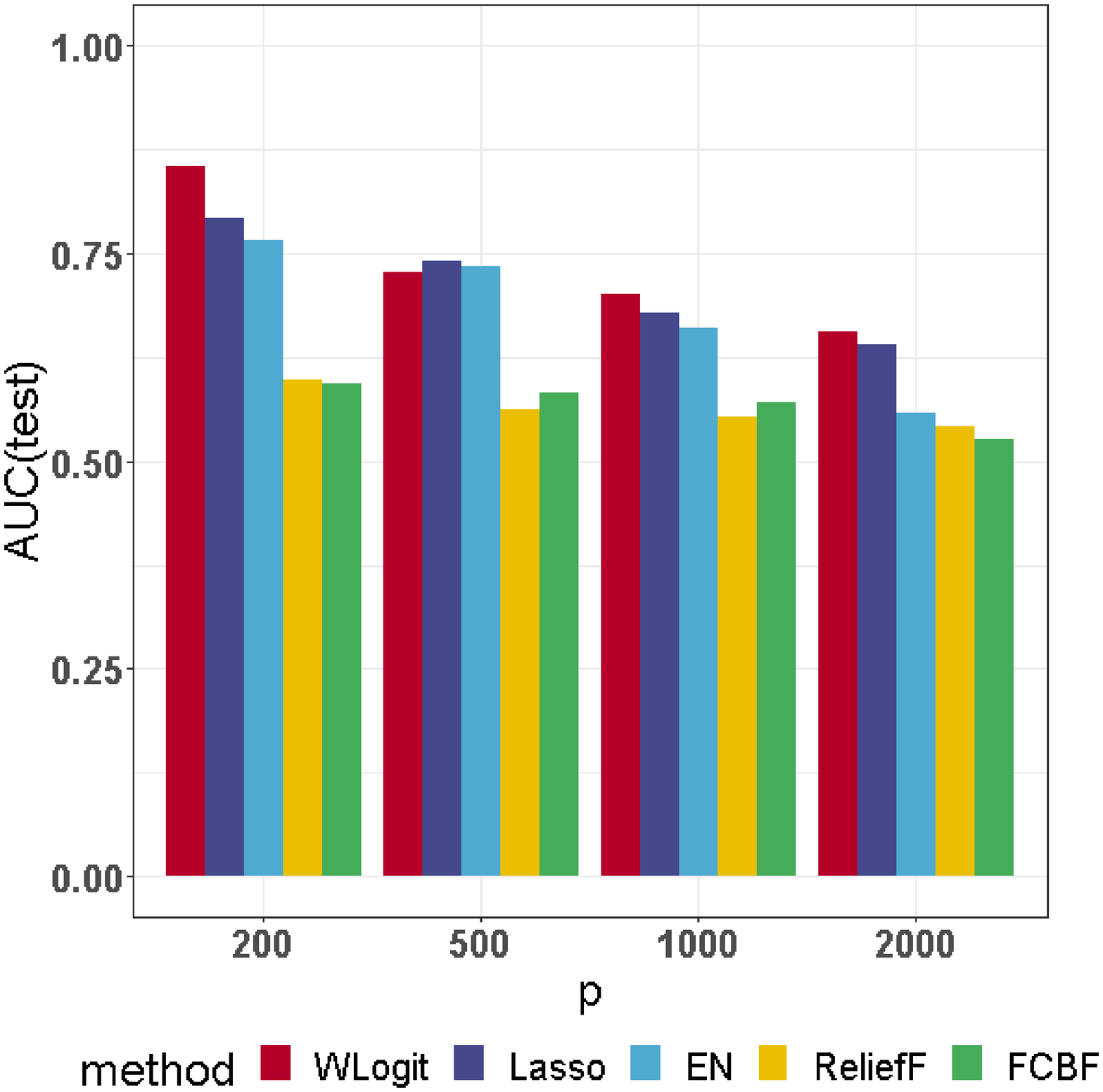}
    \caption{AUC on the testing set for different methods in the balanced case when $\bSigma$ is the identity matrix.}
    \label{fig:AUC_000_1}
\end{figure}

Figure \ref{fig:comp_TF_000_1} displays the performance of the different approaches in the case where $\bSigma=I_{p}$, when there is no correlation between the biomarkers. Even if WLogit is designed for handling the correlations when the IC is violated, it still outperformed other methods in terms of biomarker selection. The TPR is the largest among all methods, while the FPR is the smallest (FPR$<0.05$). For example when $p=2000$, the TPRs were 0.43 (WLogit), 0.25 (Lasso), 0.16 (Elastic Net), 0.03 (Relief) and 0.08 (FCBF). The FPRs for all the methods were limited. The most performant methods were then: WLogit, Lasso, Elastic Net, FCBF, and Relief, in this order. The same conclusion can be reached in sample classification accuracy from Figure \ref{fig:AUC_000_1}: WLogit always had the highest AUC (0.86 when $p=200$ and 0.66 when $p=2000$) compared to other methods. We found that a high accuracy on sample classification is usually given by a high accuracy on predictor selection.\\   

Similar results for the imbalanced case were observed and can be found in Supplementary materials. We noticed that the classification accuracy is slightly lower for all methods compared with balanced cases. However, WLogit always gives the best accuracy on both biomarker selection accuracy and sample classification.

\section{Application to gene expression data in patients with lymphoma}
\begin{figure}[h]
    \centering
    \includegraphics[width=0.5\textwidth]{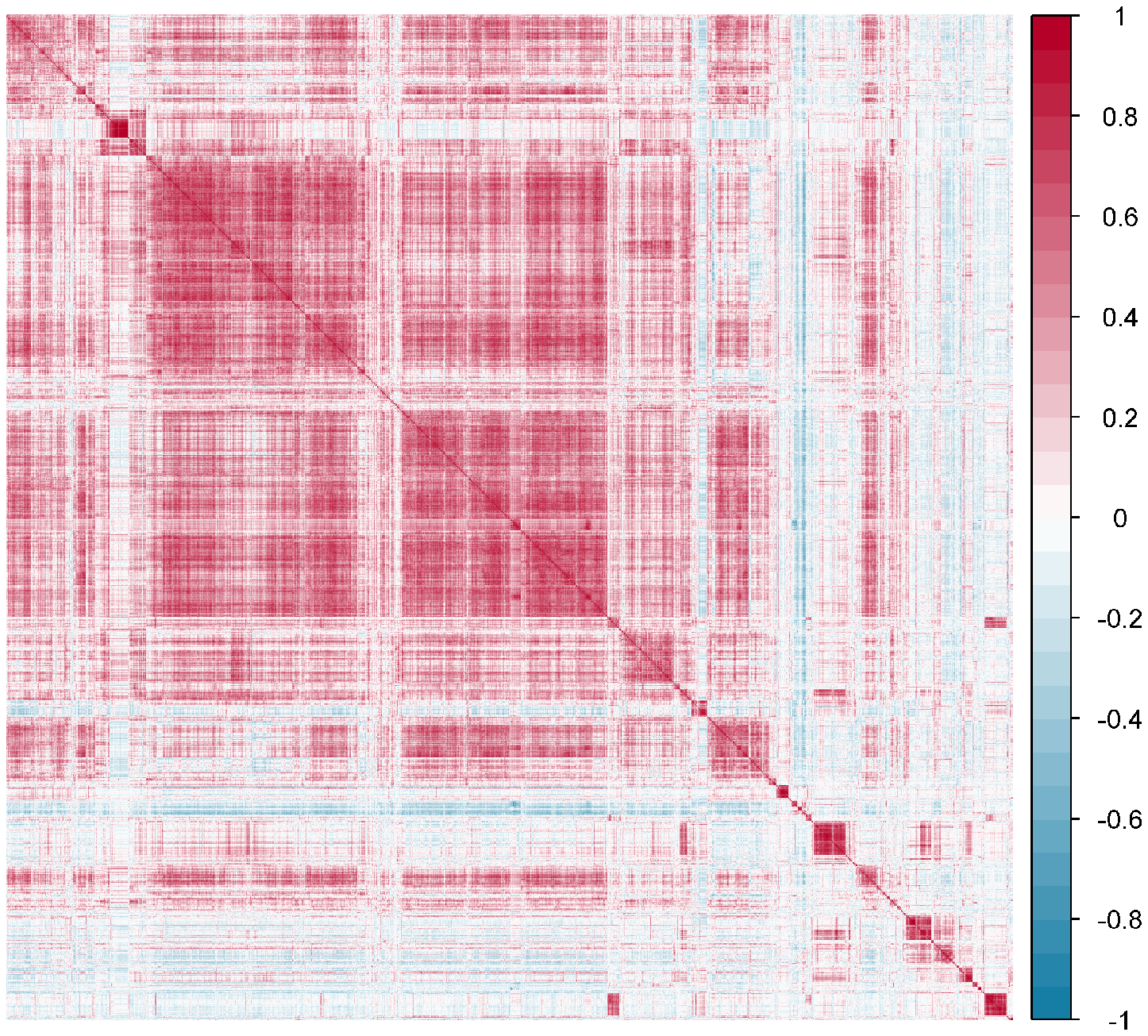}
    \caption{Heatmap of correlation of the expression of the genes in the DLBCL dataset.}
    \label{fig:corr_app}
\end{figure}

We applied the previously described approaches to gene expression data from 77 patients with lymphoma first published by \cite{Shipp2002}. This dataset contains 58 diffuse large B-cell lymphomas (DLBCL) and 19 follicular lymphomas (FL) samples. The original data contains 7,129 gene expression data. We followed the preprocessing procedures implemented in \cite{Glaab2012} which kept a total of 2648 predictors. The heatmap of the correlations between the expression of the selected genes is displayed in Figure \ref{fig:corr_app}, where we can observe strong correlations. \\

We applied different methods to select the genes that distinguish the two lymphoma subtypes (DLBCL v.s. FL). To evaluate the prediction performance of each method, we applied the commonly used 10-fold cross-validation. The dataset was separated into ten folds, and for each fit, the variable selection was conducted on the training set consisting of 90\% of the whole set. Then, the classifier was built with the subset of selected variables and used for predicting the lymphoma subtype for the remaining 10\% samples in the testing set. Finally, we report the ROC curve on the validation set and the corresponding AUC (Figure \ref{fig:AUC_app} in the Supplementary material). Our method, WLogit, achieved the highest AUC (0.95), followed by FCBF (0.85) and Relief (0.84). Lasso (0.80) and Elastic Net (0.80) both have a lower AUC; this result can come from selection failure (no predictor selected) in some folds, which degraded the overall prediction accuracy.\\

\begin{figure}[h]
    \centering
    \includegraphics[width=0.6\textwidth]{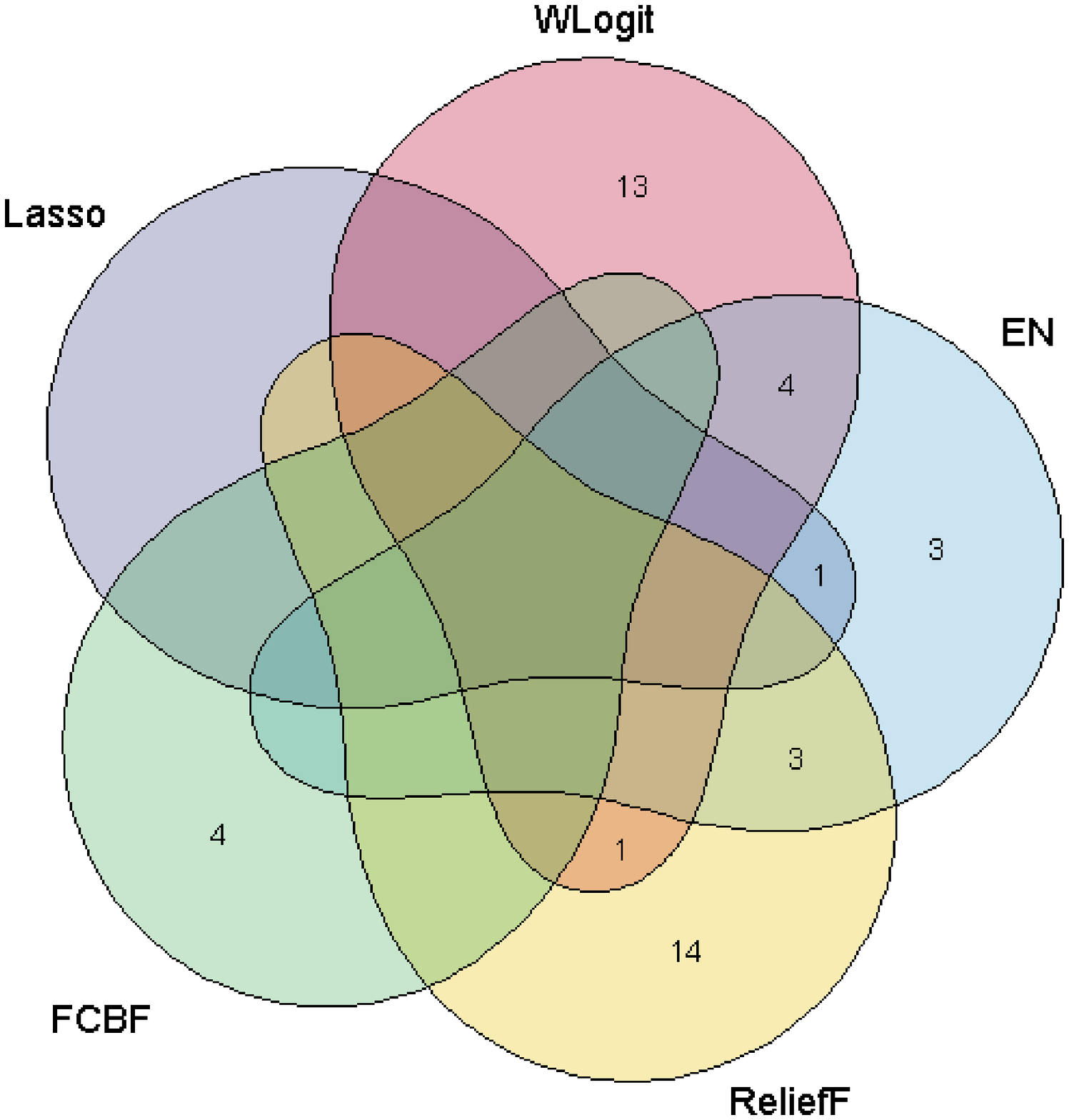}
    \caption{Venn plot of selected genes by the different compared methods.}
    \label{fig:Venn}
\end{figure}

Finally, we used the complete dataset to perform gene selection. Figure \ref{fig:Venn} presents the number of genes selected by each method and the overlap between them. WLogit selected a subset of 18 genes with four genes in common with Elastic Net and one in common with Relief. Lasso selected only one gene that was included in the set of 11 genes selected by Elastic Net. FCBF selected four genes that have no intersection with others. The list of genes selected by each method is given in Supplementary materials, with annotations provided by DAVID database \citep{Sherman2022}. 

\newpage
\section{Conclusion}
This paper proposes a novel biomarker selection method in the high dimensional logistic regression model when the biomarkers are highly correlated. Our approach, called WLogit, consists in using a penalized criterion dedicated to the logistic regression model after having removed the correlations existing between the biomarkers. The numerical experiments showed the strength of our method not only on biomarker selection but also on sample classification.\\


\clearpage

\newpage

\section*{Supplementary material}

This supplementary material provides additional numerical experiments, figures and tables for the paper: ``Variable selection in high-dimensional logistic regression models using a whitening approach''.

\begin{figure}[h]
    \centering
    \includegraphics[width=0.9\textwidth]{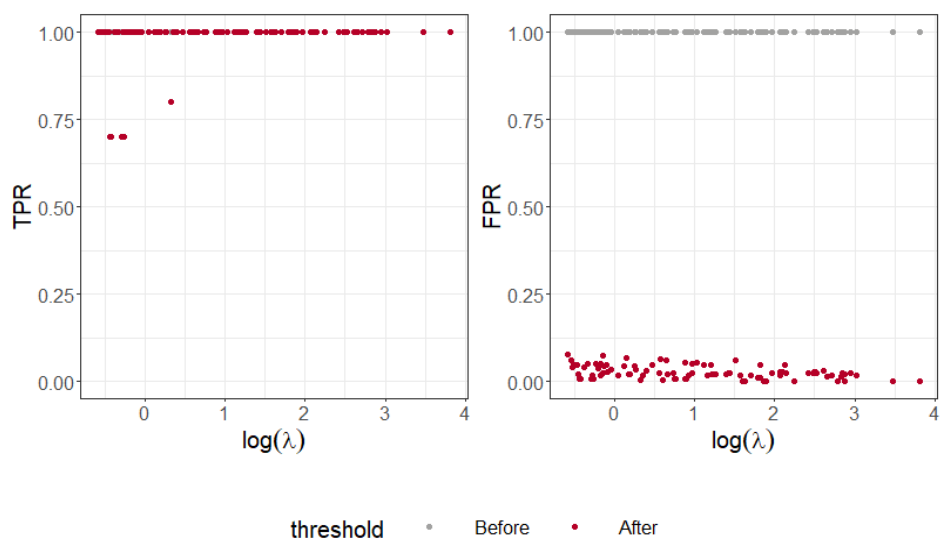}
    \caption{True Positive Rate (left) and False Positive Rate (right) before and after the thresholding.}
    \label{fig:thresh_final}
\end{figure}

\begin{figure}[h]
    \centering
    \includegraphics[width=0.9\textwidth]{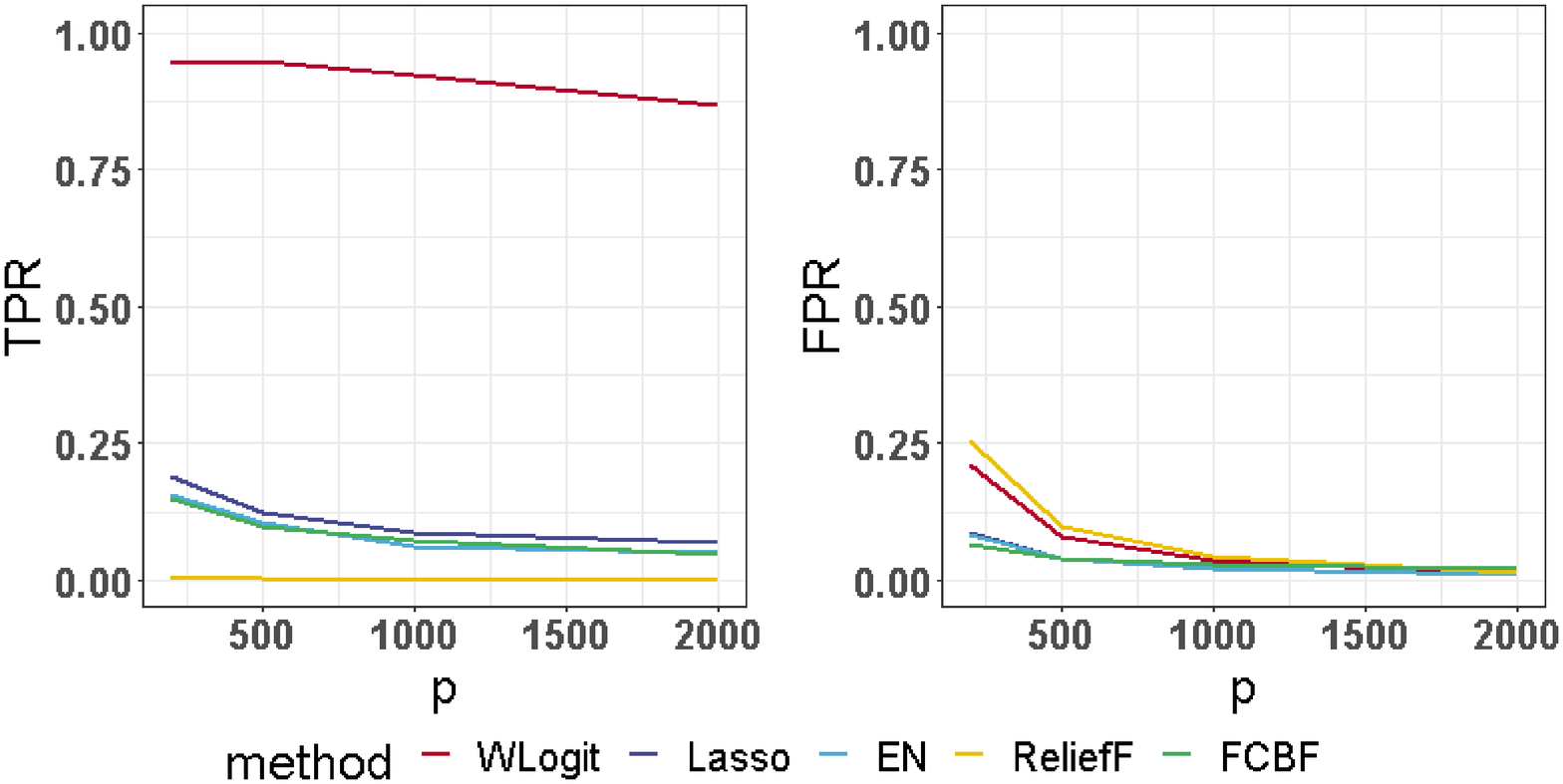}
    \caption{True Positive Rate (left) and False Positive Rate (right) for different methods in the imbalanced case when $\bSigma$ is defined in (\ref{Sigma_structure}) with $(\alpha_1,\alpha_2, \alpha_3)= (0.3, 0.5, 0.7)$.}
    \label{fig:comp_TF_357_im}
\end{figure}

\begin{figure}[h]
    \centering
    \includegraphics[width=0.9\textwidth]{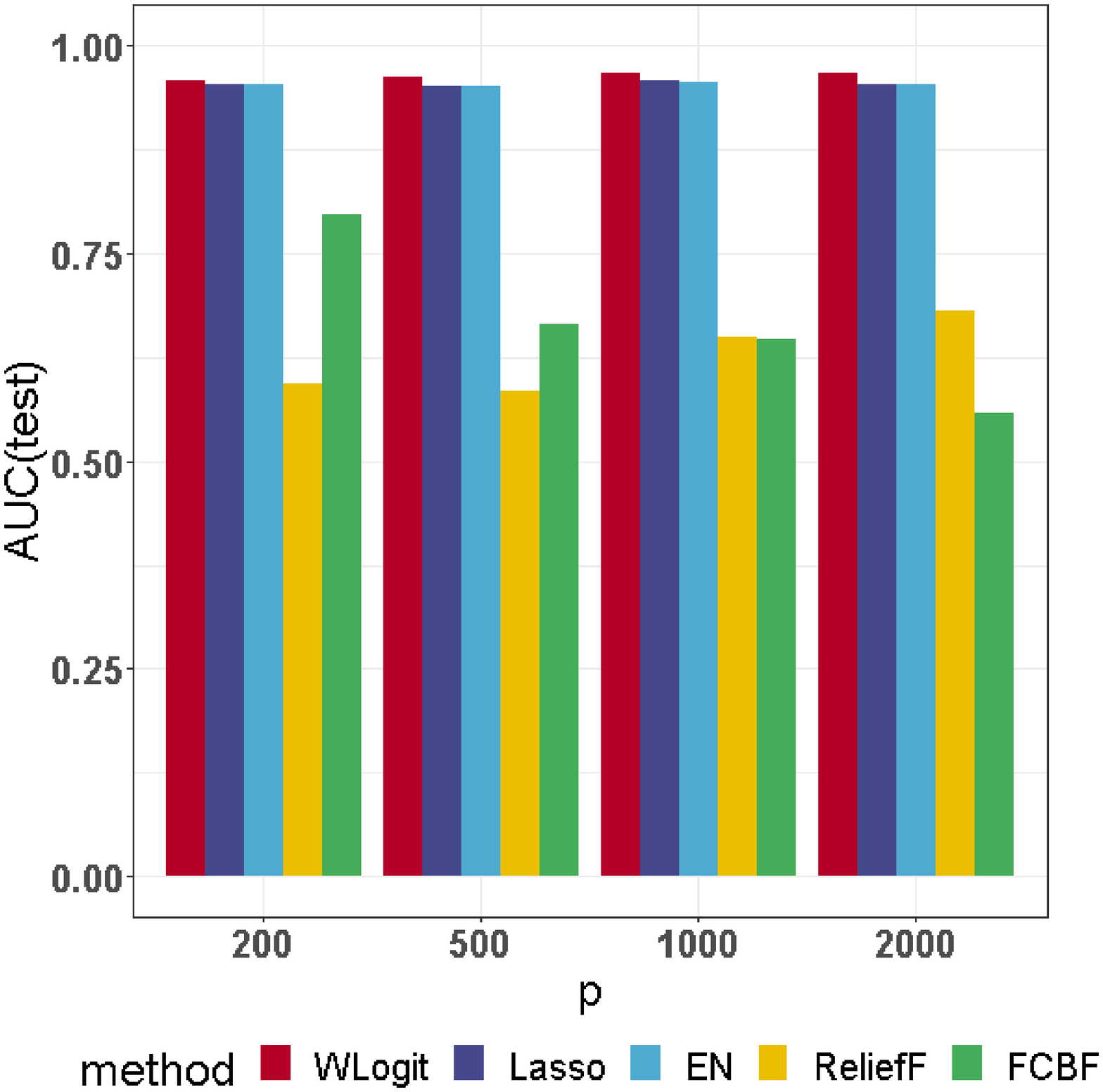}
    \caption{AUC on the testing set for different methods in the imbalanced case when $\bSigma$ is defined in (\ref{Sigma_structure}) with $(\alpha_1,\alpha_2, \alpha_3)= (0.3, 0.5, 0.7)$.}
    \label{fig:AUC_357_im}
\end{figure}

\begin{figure}[h]
    \centering
    \includegraphics[width=0.9\textwidth]{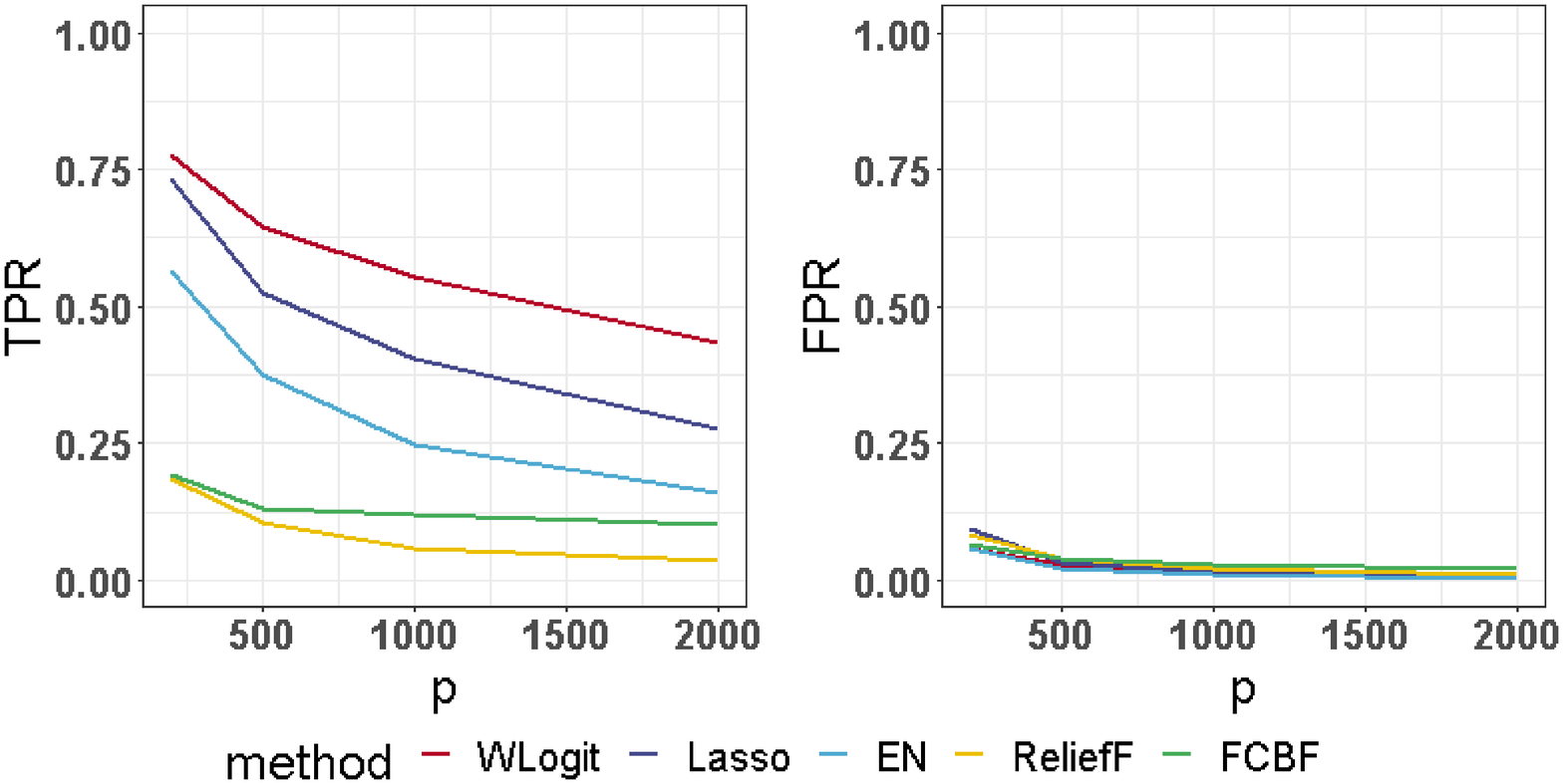}
    \caption{True Positive Rate (left) and False Positive Rate (right) for different methods in the imbalanced case when $\bSigma$ is the identity matrix.}
    \label{fig:comp_TF_000_im}
\end{figure}

\begin{figure}[h]
    \centering
    \includegraphics[width=0.9\textwidth]{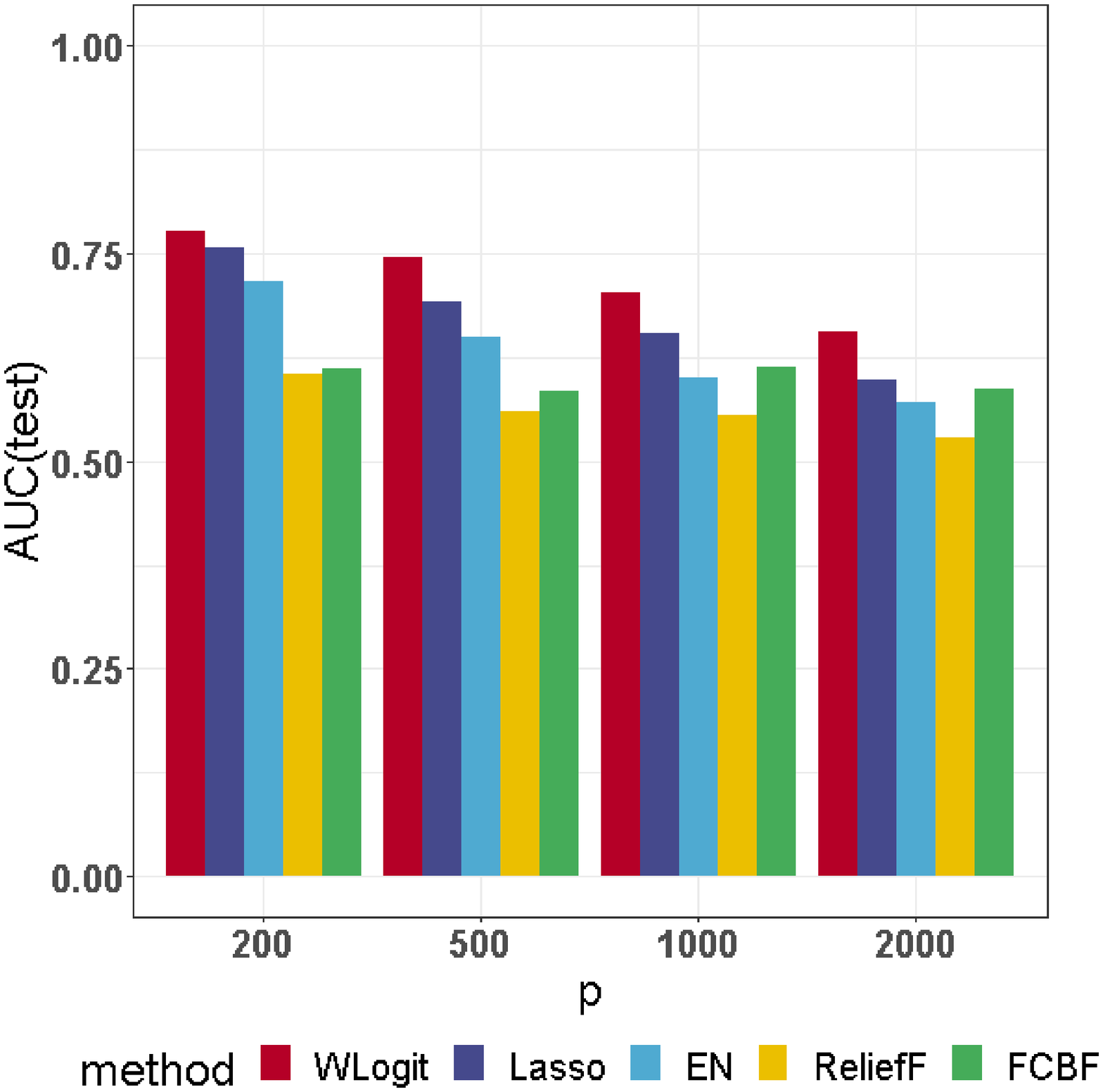}
    \caption{AUC on the testing set for different methods in the imbalanced case when $\bSigma$ is the identity matrix.}
    \label{fig:AUC_000_im}
\end{figure}

\newpage
\begin{table}[]  \caption{Selected genes and their annotations. \label{Tab:annot_genes}}
\small
\begin{tabular}{cll}
\hline
                             & \multicolumn{2}{l}{Selected genes}                                                 \\ \cline{2-3} 
\multicolumn{1}{l}{}         & ID                  & Annotation                                                   \\ \hline
\multirow{15}{*}{WLogit}     & X52773\_AT          & retinoid X receptor alpha(RXRA)                              \\
                             & D14662\_AT          & peroxiredoxin 6(PRDX6)                                       \\
                             & \textcolor{blue1}{V00594\_S\_AT}       & \textcolor{blue1}{metallothionein 2A(MT2A)}                                     \\
                             & \textcolor{blue1}{ L19686\_RNA1\_AT}    & \textcolor{blue1}{macrophage migration inhibitory factor(MIF)}                  \\
                             & AF000562\_AT        & uroplakin 2(UPK2)                                            \\
                             & D87119\_AT          & tribbles pseudokinase 2(TRIB2)                               \\
                           &  \textcolor{blue1}{ S73591\_AT }         & \textcolor{blue1}{thioredoxin interacting protein(TXNIP)}                    \\
                             & \textcolor{blue1}{X91911\_S\_AT}       & \textcolor{blue1}{GLI pathogenesis related 1(GLIPR1)}                           \\
                             & M96684\_AT          & purine rich element binding protein A(PURA)                  \\
                             & M64925\_AT          & MAGUK p55 scaffold protein 1(MPP1)                           \\
                             & U49835\_S\_AT       & chitinase 3 like 2(CHI3L2)                                   \\
                             & U14187\_AT          & ephrin A3(EFNA3)                                             \\
                             & \textcolor{blue1}{U63743\_at}        & \textcolor{blue1}{kinesin family member 2C(KIF2C)  }                            \\
                             & M63379\_AT          & clusterin(CLU)                                               \\
                             & U36787\_AT          & holocytochrome c synthase(HCCS)                              \\
   
                             & M27093\_S\_AT       & dihydrolipoamide branched chain transacylase E2(DBT)         \\ \hline
Lasso                        & \textcolor{blue1}{U63743\_at}          & \textcolor{blue1}{kinesin family member 2C(KIF2C)  }                            \\ \hline
\multirow{5}{*}{Elastic Net} & \textcolor{blue1}{AB002409\_at}        & \textcolor{blue1}{C-C motif chemokine ligand 21(CCL21)}                         \\
                             & M23323\_s\_at       & CD3 epsilon subunit of T-cell receptor complex(CD3E)         \\
                             & \textcolor{blue1}{U63743\_at}          & \textcolor{blue1}{kinesin family member 2C(KIF2C)}                              \\
                             & \textcolor{blue1}{V00594\_s\_at}       & \textcolor{blue1}{metallothionein 2A(MT2A)}                                     \\
                             & X02152\_at          & lactate dehydrogenase A(LDHA)                                \\
\multicolumn{1}{l}{}         & \textcolor{blue1}{D79987\_at}          & \textcolor{blue1}{extra spindle pole bodies like 1, separase(ESPL1)}            \\
\multicolumn{1}{l}{}         & \textcolor{blue1}{L19686\_rna1\_at}    & \textcolor{blue1}{macrophage migration inhibitory factor(MIF)}                  \\
\multicolumn{1}{l}{}        & \textcolor{blue1}{ S73591\_at}          & \textcolor{blue1}{thioredoxin interacting protein(TXNIP)}                       \\
\multicolumn{1}{l}{}         & \textcolor{blue1}{U19495\_s\_at}       & \textcolor{blue1}{C-X-C motif chemokine ligand 12(CXCL12)}                      \\ \hline
ReliefF                       & \textcolor{blue1}{AB002409\_at}        & \textcolor{blue1}{C-C motif chemokine ligand 21(CCL21)}                         \\
\multicolumn{1}{l}{}         & \textcolor{blue1}{D79987\_at}          & \textcolor{blue1}{extra spindle pole bodies like 1, separase(ESPL1)}            \\
\multicolumn{1}{l}{}         & J04031\_at          & methylenetetrahydrofolate dehydrogenase, cycslohydrolase     \\
\multicolumn{1}{l}{}         &                     & and formyltetrahydrofolate synthetase 1(MTHFD1)              \\
\multicolumn{1}{l}{}         & L00022\_s\_at       & immunoglobulin heavy constant epsilon(IGHE)                  \\
\multicolumn{1}{l}{}         & L42324\_at          & G protein-coupled receptor 18(GPR18)                         \\
\multicolumn{1}{l}{}         & M12963\_s\_at       & alcohol dehydrogenase 1A (class I), alpha polypeptide(ADH1A) \\
\multicolumn{1}{l}{}         & M15059\_at          & Fc epsilon receptor II(FCER2)                                \\
\multicolumn{1}{l}{}         & M18255\_cds2\_s\_at & protein kinase C beta(PRKCB)                                 \\
\multicolumn{1}{l}{}         & M64174\_at          & Janus kinase 1(JAK1)                                         \\
\multicolumn{1}{l}{}         & M91196\_at          & interferon regulatory factor 8(IRF8)                         \\
\multicolumn{1}{l}{}         & \textcolor{blue1}{U19495\_s\_at}       & \textcolor{blue1}{C-X-C motif chemokine ligand 12(CXCL12)}                      \\
\multicolumn{1}{l}{}         & V00594\_at          & metallothionein 1G(MT1G)                                     \\
\multicolumn{1}{l}{}         & X01677\_f\_at       & glyceraldehyde-3-phosphate dehydrogenase(GAPDH)              \\
\multicolumn{1}{l}{}         & X52142\_at          & CTP synthase 1(CTPS1)                                        \\
\multicolumn{1}{l}{}         & X69433\_at          & isocitrate dehydrogenase (NADP(+)) 2(IDH2)                   \\
\multicolumn{1}{l}{}         & \textcolor{blue1}{X91911\_s\_at}       & \textcolor{blue1}{GLI pathogenesis related 1(GLIPR1)}                           \\
\multicolumn{1}{l}{}         & Z11793\_at          & selenoprotein P(SELENOP)                                     \\ \hline
\multirow{4}{*}{FCBF}        & K02777\_s\_at       & T cell receptor delta variable 2(TRDV2)                      \\
                             & M27504\_s\_at       & DNA topoisomerase II beta(TOP2B)                             \\
                             & X52851\_rna1\_at    & peptidylprolyl isomerase A(PPIA)                             \\
                             & X67235\_s\_at       & hematopoietically expressed homeobox(HHEX)                   \\ \hline
\end{tabular}
\end{table}

\begin{figure}[h]
    \centering
    \includegraphics[width=0.9\textwidth]{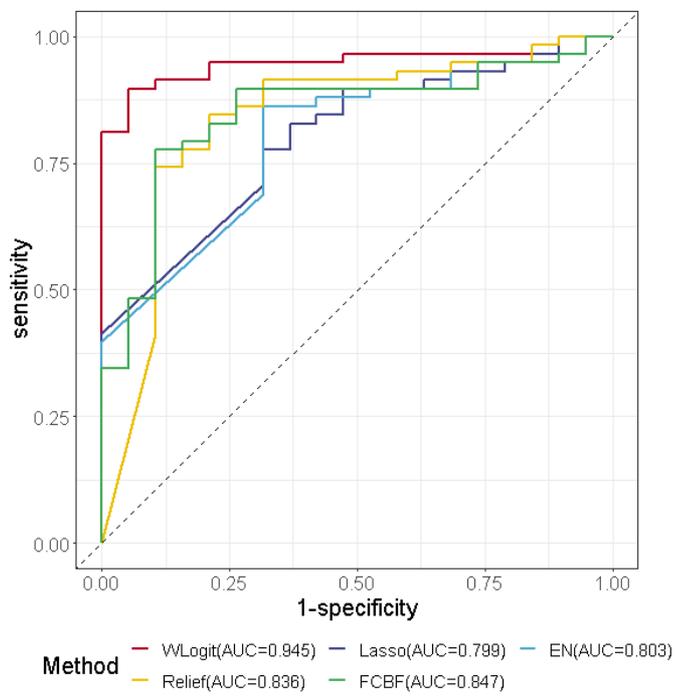}
    \caption{ROC curves and AUC for the different compared methods.}
    \label{fig:AUC_app}
\end{figure}

\end{document}